\documentclass[sigconf]{acmart}

\usepackage{makecell}
\usepackage{bm}
\usepackage{enumitem}
\setlist{nolistsep}
\usepackage{multirow}
\usepackage{float}
\usepackage{tabularx}
\usepackage{pifont}
\usepackage{booktabs}
\usepackage{caption}
\usepackage{colortbl}
\usepackage[subrefformat=parens,skip=0pt]{subcaption}
\usepackage{marginnote}
\usepackage[detect-all]{siunitx}
\usepackage{xspace}
\usepackage[bb=boondox,bbscaled=.95,cal=boondoxo]{mathalfa} 

\newcommand\ie{i.\,e.\xspace}
\newcommand\eg{e.\,g.\xspace}

\newcommand{\X}{$\mathbb{X}$\xspace}

\newcommand{\var}[1]{\mathit{#1}}

\def\sym#1{\ifmmode^{#1}\else\(^{#1}\)\fi}

\AtBeginDocument{%
  }

\copyrightyear{2026}
\acmYear{2026}
\setcopyright{cc}
\setcctype{by}
\acmConference[WWW '26]{Proceedings of the ACM Web Conference 2026}{April 13--17, 2026}{Dubai, United Arab Emirates}
\acmBooktitle{Proceedings of the ACM Web Conference 2026 (WWW '26), April 13--17, 2026, Dubai, United Arab Emirates}
\acmPrice{}
\acmDOI{10.1145/3774904.3792987}
\acmISBN{979-8-4007-2307-0/2026/04}

\begin{document}

\title{Consensus Stability of Community Notes on X}

\author{Yuwei Chuai}
\orcid{0000-0001-6181-7311}
\affiliation{
  \institution{University of Luxembourg}
  \city{Luxembourg}
  \country{Luxembourg}}
\email{yuwei.chuai@uni.lu}

\author{Gabriele Lenzini}
\orcid{0000-0001-8229-3270}
\affiliation{
  \institution{University of Luxembourg}
  \city{Luxembourg}
  \country{Luxembourg}}
\email{gabriele.lenzini@uni.lu}

\author{Nicolas Pr{\"o}llochs}
\orcid{0000-0002-1835-7302}
\affiliation{
 \institution{JLU Giessen}
 \city{Giessen}
 \country{Germany}}
\email{nicolas.proellochs@wi.jlug.de}


\begin{abstract}
Community-based fact-checking systems, such as Community Notes on \X (formerly Twitter), aim to mitigate online misinformation by surfacing annotations judged helpful by contributors with diverse viewpoints. While prior work has shown that the platform's bridging-based algorithm effectively selects helpful notes at the time of display, little is known about how evaluations change after notes become visible. Using a large-scale dataset of \num{437396} community notes and 35 million ratings from over \num{580000} contributors, we examine the stability of helpful notes and the rating dynamics that follow their initial display. We find that 30.2\% of displayed notes later lose their helpful status and disappear. Using interrupted time series models, we further show that note display triggers a sharp increase in rating volume and a significant shift in rating leaning, but these effects differ across rater groups. Contributors with viewpoints similar to note authors tend to increase supportive ratings, while dissimilar contributors increase negative ratings, producing systematic post-display polarization. Counterfactual analyses suggest that this post-display polarization, particularly from dissimilar raters, plays a substantial role in note disappearance. These findings highlight the vulnerability of consensus-based fact-checking systems to polarized rating behavior and suggest pathways for improving their resilience.
\end{abstract}



\begin{teaserfigure}
  \centering
  \includegraphics[width=\textwidth]{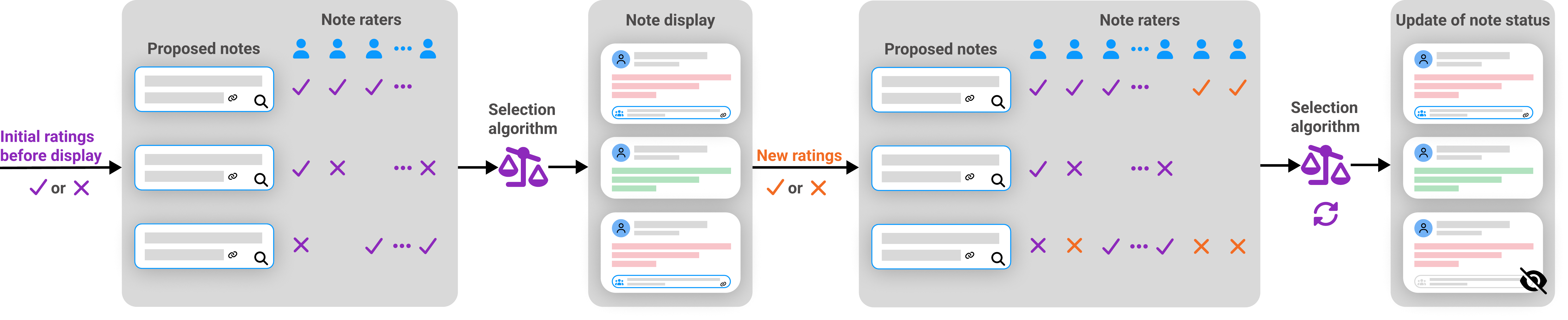}
  \caption{Workflow of the note selection algorithm. Community notes are first proposed and rated by contributors, and those deemed helpful are then publicly displayed to all users. The algorithm then continues to update note status as additional ratings are received, meaning displayed notes can later lose or regain their helpful status.}
  \label{fig:note_disappear_framework}
  \Description{}
\end{teaserfigure}


\maketitle
\section{Introduction}

The proliferation of misinformation on social media platforms poses significant risks to democratic governance, public trust, and social cohesion~\cite{ecker2024misinformation,chuai2022anger,ecker2022psychological}. In response, social media providers face mounting pressure to combat the spread of misinformation on their platforms~\cite{donovan2020social}. An increasingly adopted strategy is to leverage the ``wisdom of crowds,'' whereby users collaboratively identify misinformation and contribute context annotations for correction through community-based (crowdsourced) fact-checking~\cite{allen2021scaling}. \X's Community Notes program has emerged as a flagship implementation of this approach and has since influenced similar initiatives on other platforms, including YouTube, Facebook, and TikTok~\cite{chuai2025request}.

Participants in the Community Notes program (\ie, \emph{contributors}) can annotate potentially misleading posts by proposing \emph{community notes}, which are then evaluated by other contributors. However, unlike expert fact-checkers who provide rigorous and authoritative assessments, crowd contributors vary in expertise, consistency, and impartiality across both note writing and rating processes~\cite{allen2022birds,saeed2022crowdsourced,kaufman2022who,draws2022effects}. This creates a central challenge for community-based fact-checking: reliably surfacing accurate and broadly helpful annotations while preventing the display of low-quality, biased, or strategically manipulated notes~\cite{augenstein2025community}. To address this challenge, \X employs a bridging-based algorithm designed to identify points of agreement across politically diverse groups and promote notes that are both factually informative and widely perceived as helpful~\cite{wojcik2022birdwatch}. Prior work has shown that this algorithm can effectively select accurate notes that receive broad cross-perspective supports~\cite{allen2024characteristics,wojcik2022birdwatch}.

While the effectiveness of the note selection algorithm in surfacing helpful notes on \X has been well documented, little is known about what happens \emph{after} a note becomes visible. As illustrated in Fig.~\ref{fig:note_disappear_framework}, the algorithm continuously re-evaluates community notes by incorporating new ratings (\ie, upvotes and downvotes) on a rolling basis. As new ratings arrive, a note deemed helpful at one moment may later lose visibility; and, even factually accurate notes can lose consensus and disappear from view~\cite{allen2022birds,augenstein2025community}. This dynamic process raises the possibility that \emph{visibility itself may reshape subsequent crowd behavior}: some contributors may reinforce a note with additional support, while others may respond with disagreement, partisan backlash, or strategic downvoting. This is further complicated by potential manipulation attempts or coordinated attacks aimed at influencing note status. Together, the dynamic rating system raise an important question about the resilience of community-based fact-checking systems: do notes that initially achieve broad consensus remain stable over time, or do post-display rating dynamics (whether organic or adversarial) erode consensus? Addressing this question is the goal of our study.

\vspace{1mm}
\noindent
\textbf{Research questions.} In this study, we examine the consensus stability of community notes to assess the resilience of the note selection algorithm on \X's Community Notes platform. Using a large-scale dataset of \num{437396} Community Notes and more than 35 million ratings from over \num{583000} contributors, we address four main research questions (RQs): 
\begin{itemize}[leftmargin=*]
    \item \textit{\textbf{RQ1}: How often do community notes that are initially rated as helpful later lose their helpful status and disappear?}
    \item \textit{\textbf{RQ2}: How is note disappearance related to characteristics of the underlying posts and their authors?}
    \item \textit{\textbf{RQ3}: How do rating volume and the balance of upvotes vs. downvotes change after a note becomes visible?}
    \item \textit{\textbf{RQ4}: Do raters with similar vs. dissimilar viewpoints evaluate the same note differently once it becomes visible?}
\end{itemize}
\vspace{1mm}
\noindent
\textbf{Methodology.} To address our RQs~1--2, we employ a logistic regression model to examine the likelihood of note disappearance. For RQs~3--4, we employ a design of interrupted time series to analyze the changes in the \emph{rating count} and \emph{rating leaning} after the initial display of community notes across different rater groups. The rating count indicates the total number of ratings submitted to a note, while the rating leaning indicates the difference between upvotes and downvotes during a certain period.

\vspace{1mm}
\noindent
\textbf{Contributions.} 
Our analysis contributes four main findings:
\begin{enumerate}[leftmargin=*,label=(\roman*)]
\item We demonstrate that 30.2\% of displayed notes later lose their helpful status and disappear.
\item Note disappearance is systematically associated with characteristics of the source post: notes attached to posts about health or political topics, as well as posts authored by high-influence users, are significantly more likely to disappear. We also observe a political asymmetry such that notes displayed on posts from left-leaning authors are more likely to disappear than those on right-leaning authors' posts.
\item Using interrupted time series models, we further show that note display triggers a sharp increase in rating volume and a significant shift in rating leaning.
\item However, these shifts differ across rater groups. Contributors with viewpoints similar to note authors tend to increasingly provide supportive ratings, while dissimilar contributors increasingly provide negative ratings, producing systematic post-display polarization. Counterfactual analyses confirm that this post-display polarization, particularly from dissimilar raters, substantially contributes to note disappearance.
\end{enumerate}
Altogether, our findings highlight the vulnerability of consensus-based fact-checking systems to polarized rating behavior and suggest pathways for improving their resilience.

\section{Background and Related Work}

\textbf{Approaches to fact-checking:}
Effectively identifying and mitigating the spread of online misinformation remains a persistent challenge~\cite{ecker2022psychological}. Traditional expert-based fact-checking approaches, though capable of producing rigorous and authoritative assessments, face fundamental limitations in scalability and timeliness, rendering them insufficient for the volume and velocity of online misinformation~\cite{chuai2024roll,allen2021scaling,saeed2022crowdsourced,godel2021moderating}. To overcome these constraints, scholars and practitioners have increasingly turned to community-based (or crowdsourced) fact-checking, which leverages the crowd to collaboratively identify and contextualize misleading content~\cite{allen2021scaling,pennycook2019fighting,epstein2020will,wojcik2022birdwatch,martel2024crowds}. A growing body of research has examined expert-based (\eg, PolitiFact) and community-based (\eg, \X's Community Notes) fact-checking systems from multiple perspectives to examine fact-check generation and evaluation~\cite{mohammadi2025birdwatch}.

\textbf{Identification of misleading content:}
Current fact-checking efforts, whether conducted by professional experts or by crowds, remain inherently human-driven. Because someone must first decide which content warrants fact-checking, the selection process is vulnerable to subjective judgment and potential bias~\cite{chuai2025political,chuai2024roll,lee2023fact,draws2022effects}. Research shows that fact-checking activity tends to surge during major societal and political events such as the COVID-19 pandemic or presidential elections~\cite{chuai2025political,lee2023fact}. Partisan asymmetries in fact-checking selection have been observed in third-party organizations such as PolitiFact and Snopes. For example, PolitiFact rates Republican statements as false at roughly three times the rate of Democratic statements~\cite{ostermeier2011selection}. Additionally, fact-checked false statements from Republicans are disproportionately those that reference Democrats, oftentimes with the expression of out-group animosity~\cite{chuai2025political,rathje2021out}. 

Similar partisan asymmetries also appear in community-based fact-checking systems~\cite{truong2025community,bouchaud2025algorithmic}. On \X's Community Notes platform, contributors are more likely to flag posts from Republicans than those from Democrats~\cite{renault2025republicans}, and they tend to write community notes on posts from counter-partisans~\cite{allen2022birds}. A recent study finds that a new feature on \X, \ie, ``Request Community Note,'' introduces further partisan divergence: contributors more often annotate posts from Republicans, whereas \emph{requestors} who disproportionately submit requests targeting Democratic posts~\cite{chuai2025request}. Collectively, these findings suggest that both expert and crowd-based fact-checking processes exhibit partisan asymmetries in fact-checking selection. However, whether these asymmetries reflect inherent selection bias or the broader misinformation landscape remains an open question~\cite{chuai2025political}. Additionally, beyond partisanship, comparative analyses indicate that community-based and professional fact-checking systems, such as Community Notes and Snopes, often target different accounts: crowd fact-checkers tend to select posts from larger accounts with higher social influence, compared to expert fact-checkers~\cite{pilarski2024community}. 

\textbf{Consensus in fact-checking:}
In traditional professional fact-checking, trained experts provide rigorous and authoritative assessments of claims, typically achieving high inter-rater agreement across fact-checkers from different organizations~\cite{lee2023fact}. Conversely, community-based fact-checking relies on the collective judgment of lay contributors. The rationale is to identify misleading content by harnessing the wisdom of crowds~\cite{woolley2010evidence}. Experimental studies support this premise: the aggregated ratings of even relatively small, politically balanced crowds can approach the accuracy of professional fact-checkers~\cite{allen2021scaling,saeed2022crowdsourced}. 

However, the fact that crowds \emph{can} produce accurate judgments does not guarantee that they \emph{will}~\cite{epstein2020will}. Several challenges may compromise the integrity of community-based fact-checking systems. These include intentional manipulation or gaming of the system~\cite{luca2016fake}, limited cognitive engagement when evaluating content~\cite{pennycook2019lazy}, and motivated reasoning that shapes how users evaluate content~\cite{kahan2017motivated}. These behaviors create opportunities for users to flag or downvote fact-checks not because they are inaccurate, but because they conflict with their ideological preferences or objectives.
These challenges are further amplified by the high levels of political polarization on social media~\cite{conover2011political,barbera2015tweeting,solovev2022hate}. For example, contributors on community notes are more likely to rate fact-checks written by counter-partisans as unhelpful~\cite{allen2022birds,yasseri2023can}. To mitigate such polarization, \X introduced a bridging-based algorithm that features community notes perceived as helpful across ideologically diverse subgroups~\cite{wojcik2022birdwatch}. Previous research has confirmed the effectiveness of the algorithm in surfacing high-quality community notes for display~\cite{wojcik2022birdwatch,allen2024characteristics}. In this context, high-quality sourcing plays a key role in fostering cross-partisan agreement and enabling the algorithm to identify notes with broad support~\cite{solovev2025references,toyoda2025understanding,borenstein2025can,toyoda2025understanding}.

\textbf{Research gap.}
Different from professional fact-checking, where expert assessments are statically displayed on the corresponding misleading posts, the Community Notes system introduces a dynamic rating mechanism in which note status is continuously re-evaluated as new ratings arrive. Although the platform's bridging-based algorithm has been shown to effectively surface helpful and broadly supported notes~\cite{wojcik2022birdwatch,allen2024characteristics}, prior work has focused almost entirely on the \emph{initial} selection of notes for display. Little is known about what happens \emph{after} a note becomes visible. Because note status is dynamic, visibility itself may reshape subsequent evaluations (Fig.~\ref{fig:note_disappear_framework}), potentially reinforcing agreement among like-minded users, provoking partisan backlash, or inviting strategic downvoting. 
Understanding these post-display rating dynamics is therefore crucial for improving algorithmic design and ensuring that community-based fact-checking systems remain both effective and resilient.

\section{Data and Methods}
\subsection{Dataset}
To examine the stability of helpful community notes after their initial display, we use a large-scale dataset collected by a previous study that investigated the effectiveness of community fact-checking in reducing the spread of misleading posts~\cite{chuai2024community.new}. The dataset comprises community notes and their associated source posts over the observation period from the public roll-out of the Community Notes program on November 11, 2022, to July 14, 2024. Specifically, it includes \num{437396} community notes that had received \num{35081488} ratings from \num{583285} raters, linked to \num{233908} posts on \X. 

Out of all notes in the dataset, only \num{43782} (10.0\%) were rated helpful after leaving the initial ``Needs More Ratings'' status and were displayed on their corresponding posts (see Suppl.~\ref{supp:note_status} for details). However, \num{13202} (30.2\%) of displayed notes disappeared in later status updates. Thus, although only a small fraction of notes are ever displayed, a substantial portion of those that did appear were later removed. This pattern underscores the importance of examining the rating dynamics following the initial display of community notes. A summary of the dataset is presented in Table~\ref{tab:data_overview}. 

\begin{table}
    \centering
    \caption{Dataset overview. Columns (1)--(3) show statistics at note, post, and rating levels for displayed notes, stable notes, and displayed notes, respectively.}
    \setlength{\tabcolsep}{3pt}
    \begin{tabular}{l*{3}{c}}
    \toprule
    &{(1)} & {(2)}& {(3)}\\
    &{Displayed notes}&{Stable}&{Disappeared}\\
    \midrule
    \#Notes&{43,782}&{30,580 (69.8\%)}&{13,202 (30.2\%)}\\
    \#Writers&{19,260}&{14,257}&{8,338}\\
    Start date&{12/11/2022}&{12/11/2022}&{12/11/2022}\\
    End date&{06/27/2024}&{06/27/2024}&{06/16/2024}\\
    \multicolumn{4}{l}{\underline{Posts with community notes}}\\
    \#Posts&{35,265}&{25,437 (72.1\%)}&{11,901 (33.7\%)}\\
    \#Authors & {13,553} & {10,627} & {5,758}\\
    Start date & {12/11/2022} & {12/11/2022} & {12/11/2022}\\
    End date & {06/10/2024} & {06/10/2024} & {06/10/2024}\\
    \multicolumn{4}{l}{\underline{Ratings to community notes}}\\
    \#Ratings&{8,014,229}&{5,363,204 (66.9\%)}&{2,651,025 (33.1\%)}\\
    \#Raters&{541,833}&{508,994}&{390,285}\\
    Start date&{12/11/2022}&{12/11/2022}&{12/11/2022}\\
    End date&{07/14/2024}&{07/14/2024}&{07/13/2024}\\
    \bottomrule
    \end{tabular}
    \label{tab:data_overview}
    \Description{}
\end{table}

\subsection{Reproduction of Note Selection Algorithm}
The note selection algorithm is open-source and released by the Community Notes team on GitHub~\cite{x2024algorithm}. It is developed based on Matrix Factorization (MF) \cite{wojcik2022birdwatch}. Specifically, all ratings are structured into a sparse note-rater matrix, each rating is predicted as:
\begin{equation}
    \hat{r}_{un}= \mu + i_{u} + i_{n} + f_{u} \cdot f_{n},
\end{equation}
where $f_{u} \cdot f_{n}$ is the dot product of rater (user) vector $f_{u}$ and note vector $f_{n}$. Additionally, $\mu$ is a global intercept term, accounting for an overall propensity of raters to rate notes ``helpful'' vs. ``not helpful''; $i_{u}$ is the rater's intercept term, accounting for a user's leniency in rating notes ``helpful''; and $i_{n}$ is the note's intercept, accounting for idiosyncratic ``helpfulness'' of the note beyond that explained by rater viewpoints and leniency, \ie, the global helpfulness score assigned to the corresponding community note. Notably, the algorithm minimize a regularized least squared error loss function to estimate model parameters:
\begin{equation}
    min_{\{i,f,\mu\}}\sum_{r_{un}} (r_{un}-\hat{r}_{un})^2 + \lambda_{i}(i_{u}^2+i_{n}^2+\mu^2) + \lambda_{f}(||f_{u}||^2+||f_{n}||^2),
\end{equation}
where the regularization on the intercept terms, \ie, $\lambda_{i}=0.15$, is five times higher than the regularization on the factor terms, \ie, $\lambda_{f}=0.03$. Therefore, the algorithm is optimized toward explaining as much variation in the ratings as possible using rater and note factors, with the high estimated $i_{n}$ indicating that the corresponding note is perceived helpful across diverse rater representations, \ie, rater factor vectors. Additionally, to avoid overfitting, the note selection algorithm uses one-dimensional rater and note factors. 

We use the version of the algorithm source code available at the time of data collection to calculate note intercepts and note factors for our analysis. The estimated \emph{Core note intercept} reflects the perceived helpfulness of a community note, while the estimated \emph{Core note factor} captures the polarization in its ratings~\cite{wojcik2022birdwatch} (see distribution and validation of these measures are provided in Suppl.~\ref{supp:note_selection_algorithm}).

\subsection{Matrix Factorization for Rater Similarity}
Given that the note selection algorithm emphasizes agreement among raters with diverse perspectives, it is essential to quantify the similarity between raters based on their past rating behaviors. Notably, the rater factor used by the Community Notes algorithm is intentionally one-dimensional to avoid overfitting on the initially small set of contributors, though higher-dimensional representations are expected as the contributor base grows~\cite{wojcik2022birdwatch}. In our study, given the large number of contributors and our focus on estimating nuanced rater representations rather than reducing overfitting, we adapt a matrix factorization algorithm that increases the dimensionality of both note and rater factors to 200. Additionally, to ensure sufficient information for model training, we restrict the dataset to notes that have received at least five ratings and raters who have rated at least five notes. The resulting dataset includes \num{34596349} ratings submitted by \num{406181} raters on \num{372980} notes. 

Since our goal is to accurately estimate latent rater representations from the observed ratings without emphasizing generalization beyond the existing dataset, we omit the regularization terms and use a mean squared error loss function to optimize model fit:
\begin{equation}
    min_{\{i,f,\mu\}} \frac{1}{R} \sum_{r_{un}} (r_{un}-\hat{r}_{un})^2.
\end{equation}
After 20 epochs of training, the loss converges to a stable level (see Fig.~\ref{fig:mseloss}, Suppl.~\ref{supp:mf}), and the model demonstrates strong performance on the observed ratings, achieving a weighted $F1$ score of 0.962, a weighted precision of 0.969, and a weighted recall of 0.955 (see the validation of the model estimates in Suppl.~\ref{supp:mf}).

\textbf{Similarity among raters.}
We calculate the similarity between raters that submitted ratings on the same note based on the 200-dimensional rater representations produced by the adapted MF model. Specifically, for each note, we examine all rater pairs and restrict our analysis to pairs that have co-rated at least five notes. we distinguish between rater pairs that assigned the same helpfulness level to a note and those that assigned opposite helpfulness levels. In total, this results in \num{78166385} rater pairs. We calculate the similarity for each rater pair and examine the distributions of similarities for rater pairs with same ratings on the same notes (see Fig.~\ref{fig:cosine_dist_agree}; mean $=$ \num{0.136}, std. $=$ \num{0.097}) and rater pairs with the opposite ratings on the same notes (see Fig.~\ref{fig:cosine_dist_disagree}; mean $=$ \num{-0.097}, std. $=$ \num{0.094}). Additionally, we observe a clear bimodal pattern in the overall distribution of rater similarities across all co-rating pairs (Fig.~\ref{fig:cosine_dist}), suggesting the presence of distinct groups of raters with different viewpoints. Therefore, in the subsequent analysis of the relationship between note writers and raters, we categorize raters as having \emph{similar viewpoints} to a note writer if their similarity scores exceed ($>$) \num{0.136}, and as having \emph{opposite viewpoints} if their similarity scores are below ($<$) \num{-0.097}, based on their previous co-rating histories. {Raters whose similarity scores to note writers lie between these thresholds are categorized as \emph{general raters}.}

\subsection{Estimation Models}
In this study, we employ a series of regression models and conduct empirical analyses based on the observational dataset to address our research questions. To address RQs~1--2, we specify a multilevel mixed-effects logistic regression model:
\begin{equation}
    logit(P(Y_{ij})) = \beta_{0} + \bm{\beta_{1}'X_{i}^{Post}} + \bm{\beta_{2}'X_{j}^{Poster}} + u_{j} + v_{i(j)},
\end{equation}
where $Y_{ij}$ indicates whether the community note for post $i$ from its author $j$ disappears after its initial display ($=$ 1) or not ($=$ 0), and $\beta_{0}$ is the intercept. The vector $\bm{X_{i}^{Post}}$ represents variables related to post $i$, and the vector $\bm{X_{j}^{Poster}}$ represents variables related to post author $j$. Details on the post characteristics and author attributes included in the dataset are provided in Suppl.~\ref{supp:post_characteristics}. Additionally, given that each post author can have multiple posts annotated by community notes, we incorporate multilevel random effects at both the post author level and post level. Specifically, $u_{j}$ captures the random effect at the post author level, and $v_{i(j)}$ indicates the random effect at the post level, which is nested within post author level.

To estimate the changes in rating count and rating leaning after the initial display of community notes and to address RQs~3--4, we adopt an interrupted time series design specified as following:
\begin{equation}
    Y_{i} = \beta_{0} + \beta_{1}T_{i} + \beta_{2}D_{i} + \beta_{3}D_{i} \times T_{i}  + \gamma_{i} + \epsilon_{i},
\end{equation}
where $Y_{i}$ represents either the \emph{rating count} or the \emph{rating leaning} for note $i$ at time $T_{i}$. Here, a positive rating leaning (upvotes $-$ downvotes) suggests that contributors prefer to rate notes as helpful, while a negative rating leaning suggest that contributors prefer to rate a note as not helpful. The variable $T_{i}$ indicates the number of 15-min intervals (referred to as ``quarters'') relative to the initial display of the community note (negative values indicate periods before display, and positive values indicate periods after display). The dummy variable $D_{i}$ is a binary indicator equal to 1 for the period after the note's initial display and 0 for the period before display. The interaction term $D_{i} \times T_{i}$ indicates the number of quarters since the initial display of the community note, while equal to 0 before display. Therefore, the estimate of $\beta_{2}$ captures the immediate effect at the display intervention, and the estimate of $\beta_{3}$ captures the sustained effect of the display intervention. Additionally, $\gamma_{i}$ accounts for note-specific fixed effects, and $\epsilon_{i}$ is the error term.

Following previous research~\cite{horta2025post}, we estimate a Poisson regression model with robust standard errors for the rating count to account for the non-negative and discrete nature of count data. For rating leaning, which is a continuous variable ranging from negative (more not-helpful ratings) to positive (more helpful ratings), we estimate an OLS linear regression model with robust standard errors. Notably, we also cluster robust standard errors at the note level to account for potential within-note correlations over time~\cite{chuai2025community}.

\section{Results}
\subsection{Note Disappearance (RQ1 \& RQ2)}
\begin{figure}
    \centering
    \includegraphics[width=\linewidth]{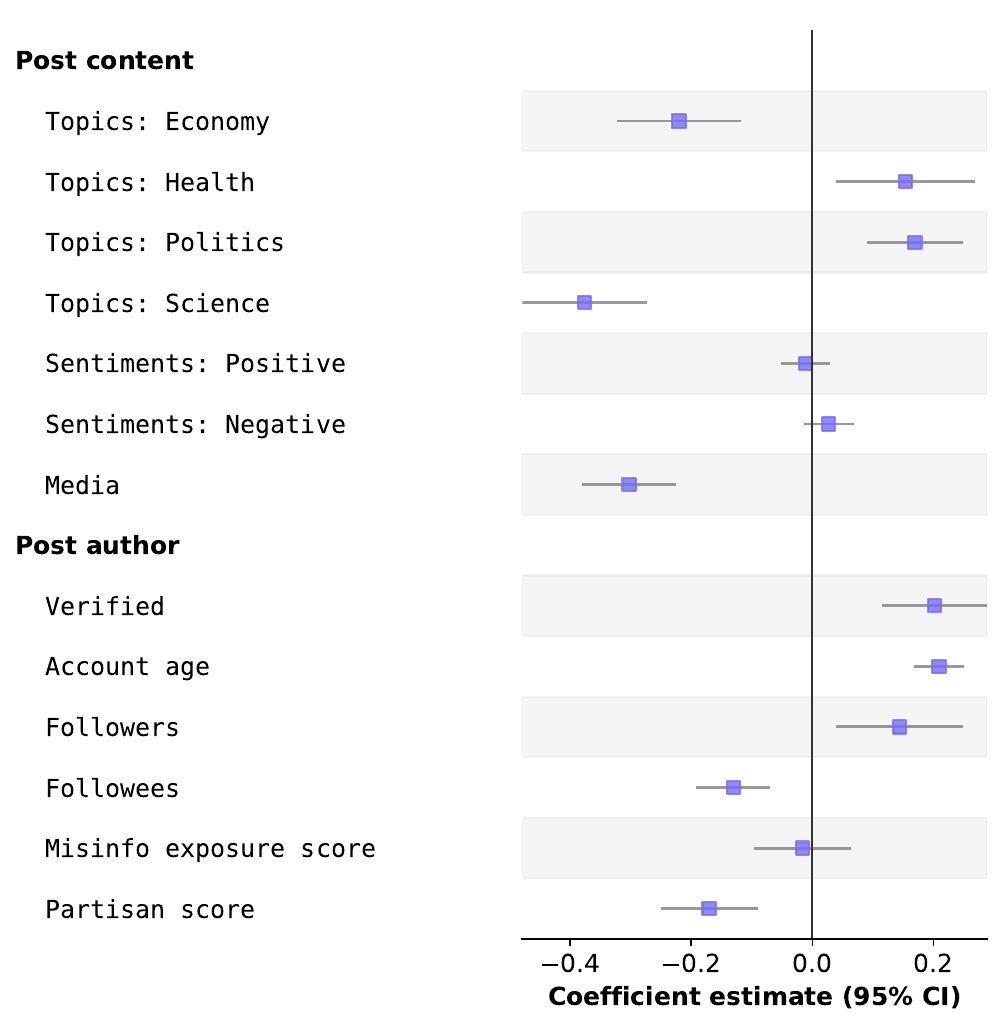}
    \caption{Estimation results for the likelihood of note disappearance. Shown are coefficient estimates with 95\% Confidence Intervals (CIs). Random effects at both the author level and the post level are included during estimation.}
    \label{fig:note_disappear_coefs}
    \Description{}
\end{figure}

We start by assessing the stability of the helpful status of community notes and the factors that are associated with their disappearance following initial display. As shown in Table~\ref{tab:data_overview}, 30.2\% (\num{13202}) of displayed community notes later disappeared, indicating a substantial rate of removal. Figure~\ref{fig:note_disappear_coefs} presents the estimated effects on the likelihood of note disappearance.

With respect to topic categories, community notes on posts about the economy ($\var{coef.}=\num{-0.220}$, $p<\num{0.001}$) or science ($\var{coef.}=\num{-0.376}$, $p<\num{0.001}$) are significantly less likely to disappear compared to notes on posts without these topics. Conversely, notes on posts related to health ($\var{coef.}=\num{0.154}$, $p<\num{0.01}$) or politics ($\var{coef.}=\num{0.169}$, $p<\num{0.001}$) are significantly more likely to disappear compared to notes on posts without corresponding topics. We find no significant relationship between note disappearance and either positive or negative sentiment in the source post; however, the presence of media elements is associated with a reduced likelihood of disappearance ($\var{coef.}=\num{-0.302}$, $p<\num{0.001}$). Author-level characteristics further shape disappearance patterns. Notes displayed on posts from high-influence users, \ie, those with verified status ($\var{coef.}=\num{0.202}$, $p<\num{0.001}$), older accounts ($\var{coef.}=\num{0.209}$, $p<\num{0.001}$), or larger numbers of followers ($\var{coef.}=\num{0.144}$, $p<\num{0.01}$), are more likely to disappear than notes on posts from low-influence users. In contrast, notes attached to posts from authors with more followees ($\var{coef.}=\num{-0.130}$, $p<\num{0.001}$) or right-leaning partisanship ($\var{coef.}=\num{-0.170}$, $p<\num{0.001}$) are less likely to disappear than those from authors with fewer followees or left-leaning partisanship.

\begin{figure*}
    \centering
    \begin{subfigure}{0.49\textwidth}
    \caption{}
    \includegraphics[width=\textwidth]{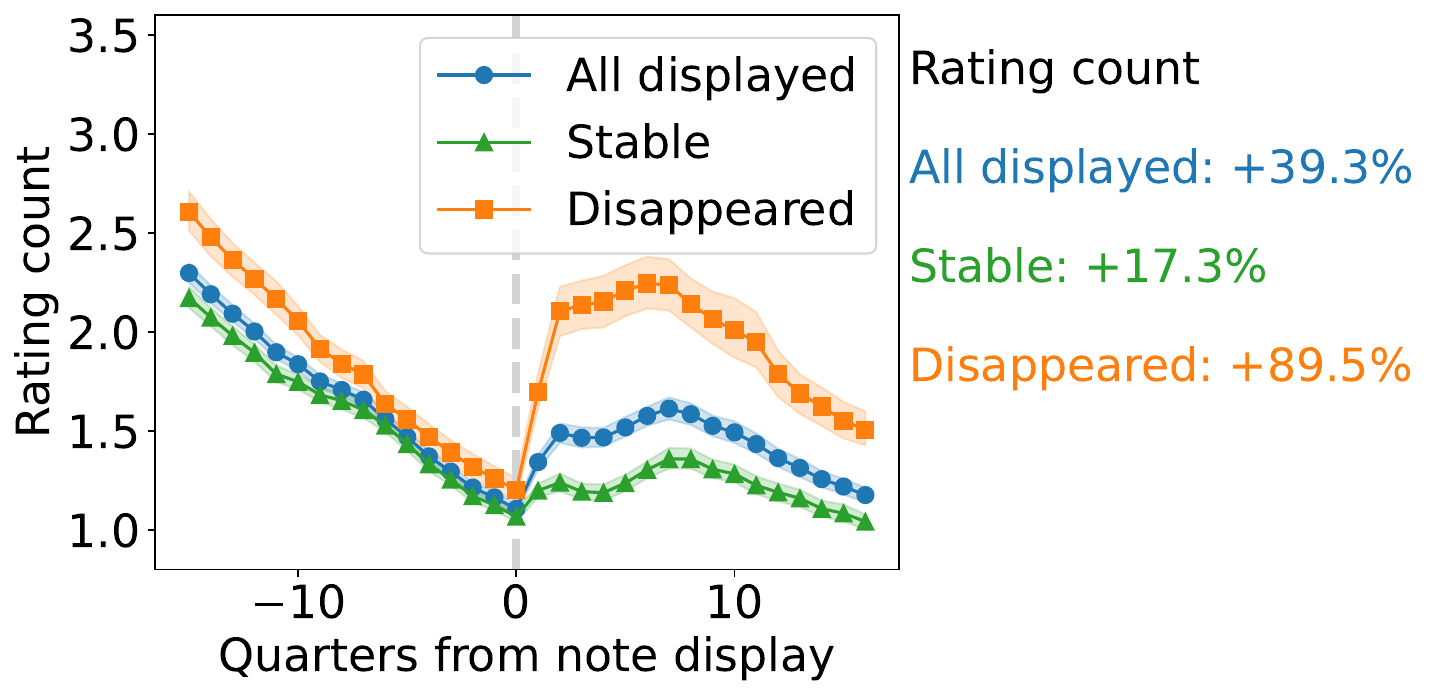}
    \label{fig:quarter_rating_count}
    \end{subfigure}
    \hfill
    \begin{subfigure}{0.49\textwidth}
    \caption{}
    \includegraphics[width=\textwidth]{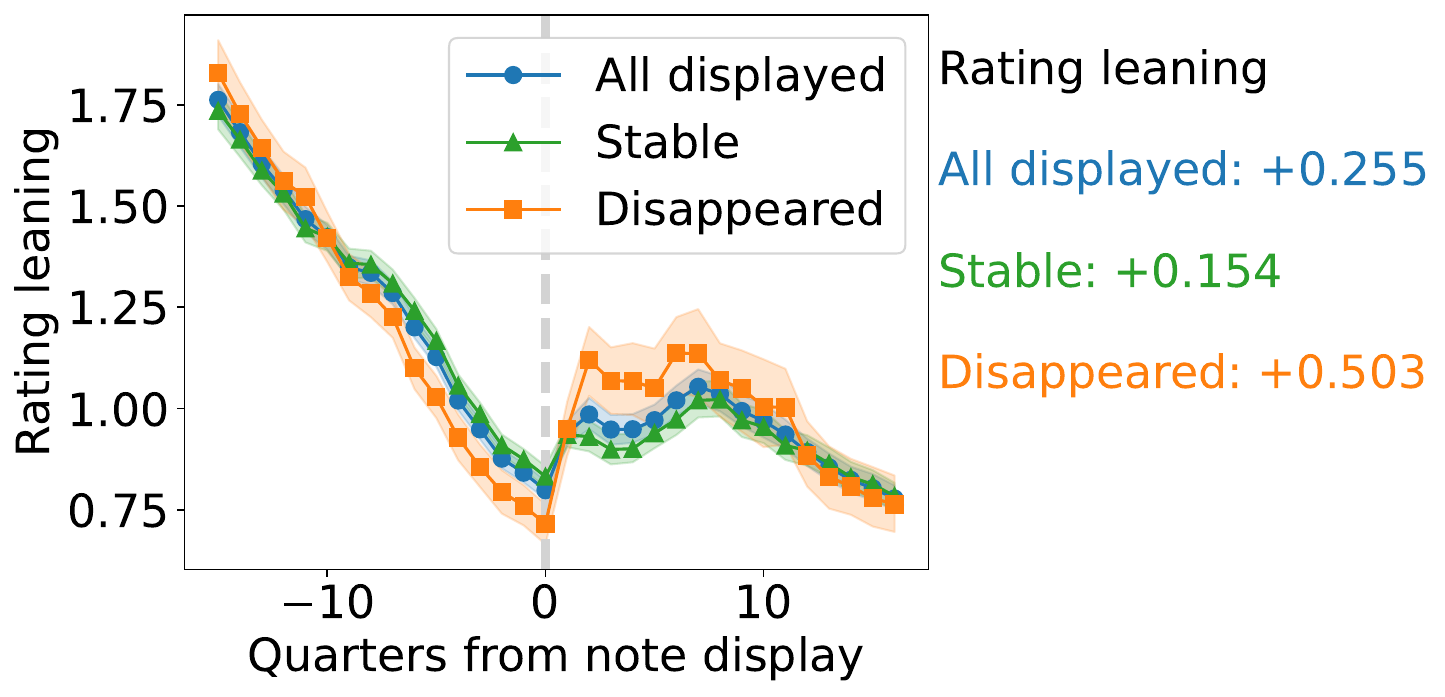}
    \label{fig:quarter_rating_leaning}
    \end{subfigure}
    \caption{Changes in rating count and rating leaning after the initial display of community notes. (a)~Average rating count per quarter (15-minute interval) from \num{-16} to 16 quarters relative to note display. (b)~Average rating leaning per quarter over the same window. The results are shown for all displayed notes, stable notes, and disappeared notes. Error bands represent 95\% CIs.}
    \label{fig:rating_count_leaning}
    \Description{}
\end{figure*}

\subsection{Changes in Rating Count and Leaning (RQ3)}

\textbf{Changes in rating count.} We examine the rating counts over an observation window spanning \num{-16} to \num{16} quarters relative to the initial display of community notes and estimated the effect of note display using an ITS design. As shown in Fig.~\ref{fig:quarter_rating_count}, note display produces a sharp immediate increase of 39.3\% (95\% CI: [35.7\%, 43.0\%]; $p<$ \num{0.001}) in the number of ratings compared to the pre-display period. Notably, this immediate effect is substantially larger for disappeared notes (89.5\%, $p<$ \num{0.001}; 95\% CI: [80.3\%, 99.2\%]) than for stable notes (17.3\%, $p<$ \num{0.001}; 95\% CI: [14.3\%, 20.4\%]). Additionally, a relatively modest yet statistically significant sustained increase of 3.6\% per quarter (95\% CI: [3.3\%, 3.9\%]; $p<$ \num{0.001}) is observed over time, which is similar across disappeared notes (3.4\%, $p<$ \num{0.001}; 95\% CI: [2.8\%, 4.0\%]) and stable notes (3.8\%, $p<$ \num{0.001}; 95\% CI: [3.5\%, 4.2\%]). Taken together, these findings indicate that the primary difference in the effect of note display between disappeared notes and stable notes lies in the magnitude of the immediate post-display response. Accordingly, in the subsequent analyses, we focus on the immediate effect following note display, which captures the direct impact of the display intervention.

\textbf{Changes in rating leaning.} 
Fig.~\ref{fig:quarter_rating_leaning} shows that the rating leaning increases by \num{0.255} (95\% CI: [\num{0.225}, \num{0.286}]; $p<$ \num{0.001}) following note display. This suggests that the increased ratings observed in Fig.~\ref{fig:quarter_rating_count} are primarily driven by upvotes exceeding downvotes. Particularly, the increase in rating leaning for disappeared notes (\num{0.503}, $p<$ \num{0.001}; 95\% CI: [\num{0.428}, \num{0.577}]) is significantly larger than that for stable notes (\num{0.154}, $p<$ \num{0.001}; 95\% CI: [\num{0.124}, \num{0.185}]). 

\textbf{Sensitivity analysis.} 
We further examine changes in rating leaning across post content categories and author characteristics (see Suppl.~\ref{supp:sensitivity}). We find that the estimated effect sizes vary across subgroups, but almost all are significantly positive (with the exception of health-related posts). This suggests that note display consistently attracts more upvotes than downvotes across a wide range of post content types and author attributes.

\begin{figure*}
    \centering
    \begin{subfigure}{0.32\textwidth}
    \caption{}
    \includegraphics[width=\textwidth]{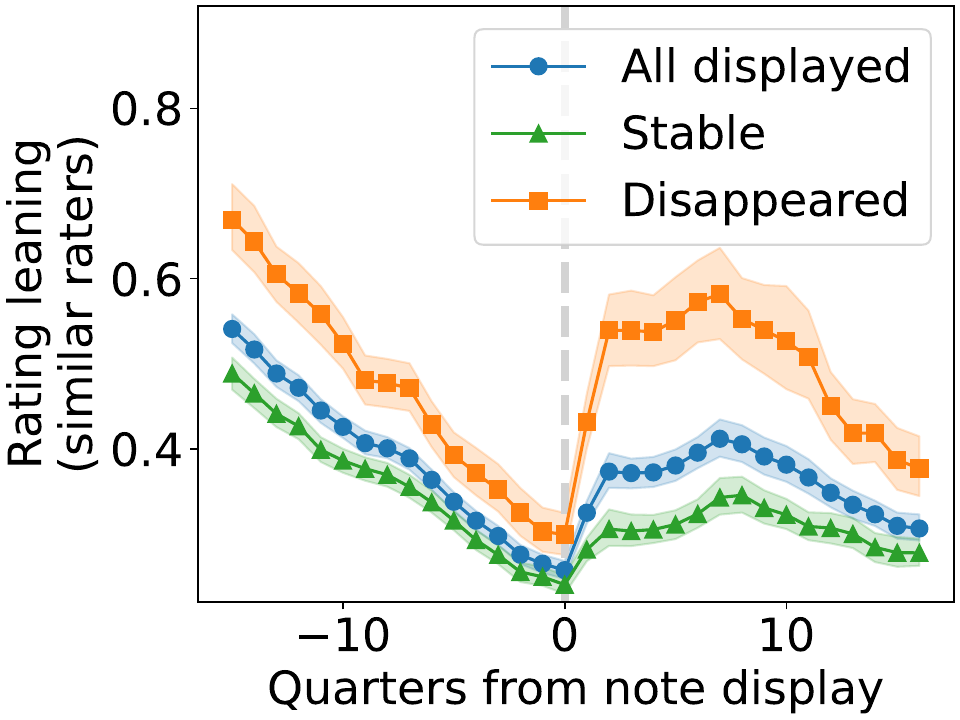}
    \label{fig:similar_rating_leaning}
    \end{subfigure}
    \hfill
    \begin{subfigure}{0.32\textwidth}
    \caption{}
    \includegraphics[width=\textwidth]{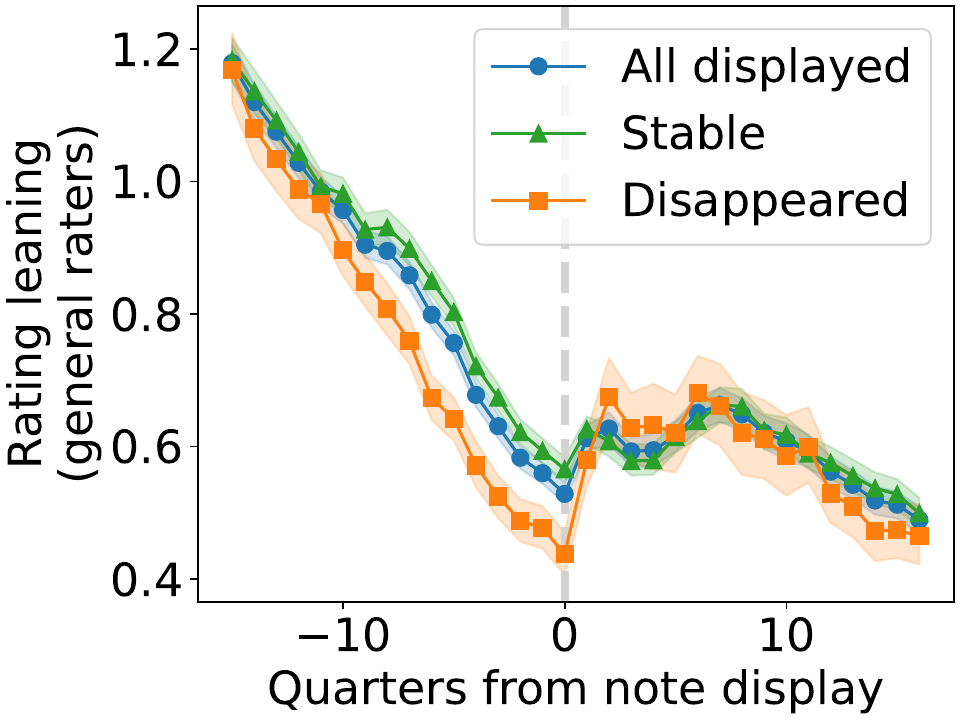}
    \label{fig:general_rating_leaning}
    \end{subfigure}
    \hfill
    \begin{subfigure}{0.32\textwidth}
    \caption{}
    \includegraphics[width=\textwidth]{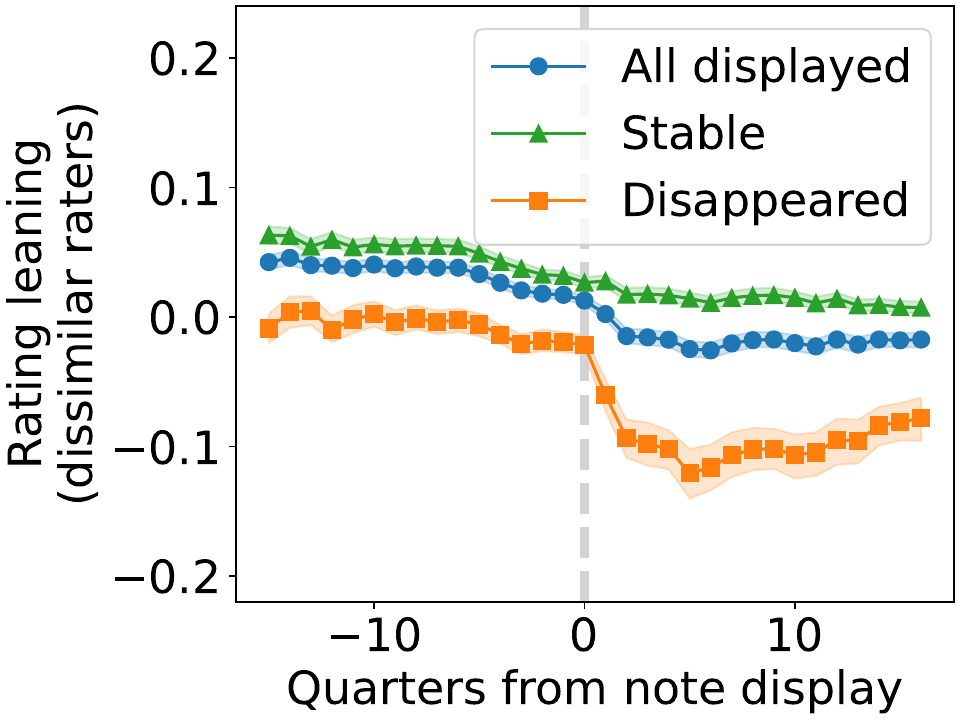}
    \label{fig:dissimilar_rating_leaning}
    \end{subfigure}
    \caption{Discrepancies in rating leaning between similar and dissimilar raters. (a)~Average rating leaning per quarter (15-minute interval) for similar raters from \num{-16} to 16 quarters relative to note display. (b)~Average rating leaning per quarter for general raters over the same window. (c)~Average rating leaning per quarter for dissimilar raters over the same window. The results are shown for all displayed notes, stable notes, and disappeared notes. Error bands represent 95\% CIs.}
    \label{fig:rating_similarity}
    \Description{}
\end{figure*}

\subsection{Similar vs. Dissimilar Raters (RQ4)}
Although community notes generally receive more upvotes than downvotes following their initial display, many still disappear. As shown earlier, disappeared notes tend to attract a greater number of ratings and exhibit higher rating leaning after display compared to stable notes. Given that the note selection algorithm emphasizes agreement among raters with diverse viewpoints, it is possible that the increased ratings following note display are biased or polarized, thereby increasing the likelihood of disappearance. To investigate this, we examine differences in rating leaning between raters whose past rating behavior is similar to that of the note writer and those who are dissimilar.

In our dataset, a total of \num{1402344} ratings fall within the observation window spanning \num{-16} to \num{16} quarters from note display, of which \num{371277} (26.5\%) originate from similar raters, and \num{110103} (7.9\%) from dissimilar raters. Thus, viewpoint-aligned and viewpoint-opposed raters constitute a minority, but potentially an influential subgroup relative to the majority of general raters (65.6\%).

Figure~\ref{fig:similar_rating_leaning} shows that similar raters exhibit a sharp increase in rating leaning following note display ($\var{coef.}=$ \num{0.148}, $p<$ \num{0.001}; 95\% CI: [\num{0.133}, \num{0.163}]). The increase in rating leaning for disappeared notes ($\var{coef.}=$ \num{0.298}, $p<$ \num{0.001}; 95\% CI: [\num{0.262}, \num{0.334}]) is significantly larger than that for stable notes ($\var{coef.}=$ \num{0.087}, $p<$ \num{0.001}; 95\% CI: [\num{0.073}, \num{0.101}]). General raters exhibit a similar pattern (Fig.~\ref{fig:general_rating_leaning}), showing a significant increase in rating leaning following note display ($\var{coef.}=$ \num{0.139}, $p<$ \num{0.001}; 95\% CI: [\num{0.119}, \num{0.158}]). Likewise, the increase for disappeared notes ($\var{coef.}=$ \num{0.287}, $p<$ \num{0.001}; 95\% CI: [\num{0.240}, \num{0.334}]) is significantly stronger than that for stable notes ($\var{coef.}=$ \num{0.070}, $p<$ \num{0.001}; 95\% CI: [\num{0.059}, \num{0.098}]). In contrast, Fig.~\ref{fig:dissimilar_rating_leaning} shows that the rating leaning for raters who have dissimilar viewpoints with note authors drops significantly following note display ($\var{coef.}=$ \num{-0.032}, $p<$ \num{0.001}; 95\% CI: [\num{-0.037}, \num{-0.027}]). This decrease is more pronounced for disappeared notes ($\var{coef.}=$ \num{-0.083}, $p<$ \num{0.001}; 95\% CI: [\num{-0.096}, \num{-0.069}]) compared to stable notes ($\var{coef.}=$ \num{-0.011}, $p<$ \num{0.001}; 95\% CI: [\num{-0.015}, \num{-0.008}]). 

Taken together, these results reveal a pattern of polarized rating behavior following note display: supportive ratings increase among similar raters while agreement decreases among dissimilar raters, with polarization substantially stronger for notes that ultimately disappear. This pattern is robust across post content categories and author characteristics, with especially pronounced polarization for political posts and left-leaning authors (see Fig.~\ref{fig:polarized_leaning_sensitivity}, Suppl.~\ref{supp:sensitivity}). 

\begin{figure*}
    \centering
    \begin{subfigure}{0.32\textwidth}
    \caption{}
    \includegraphics[width=\textwidth]{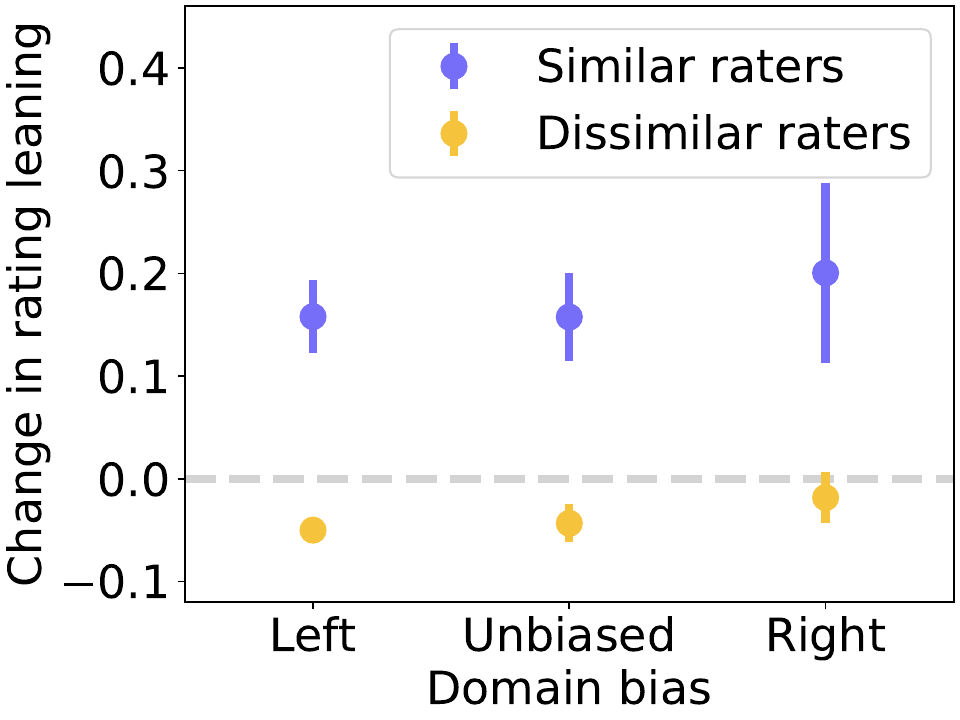}
    \label{fig:domain_bias_rating_leaning}
    \end{subfigure}
    \hspace{10em}
    \begin{subfigure}{0.32\textwidth}
    \caption{}
    \includegraphics[width=\textwidth]{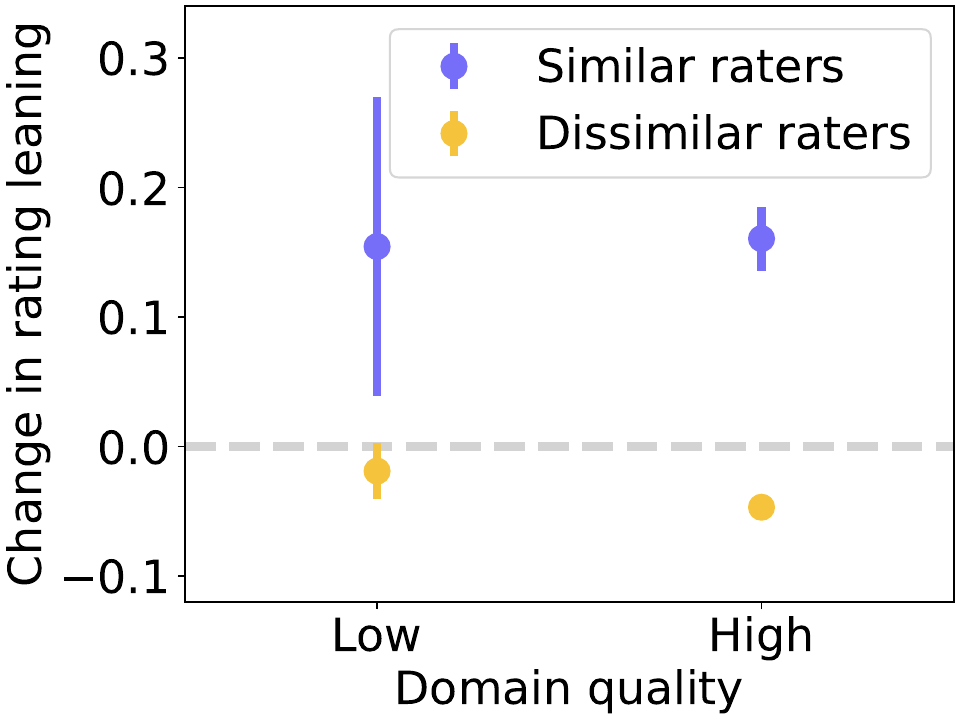}
    \label{fig:domain_quality_rating_leaning}
    \end{subfigure}
    
    \begin{subfigure}{0.32\textwidth}
    \caption{}
    \includegraphics[width=\textwidth]{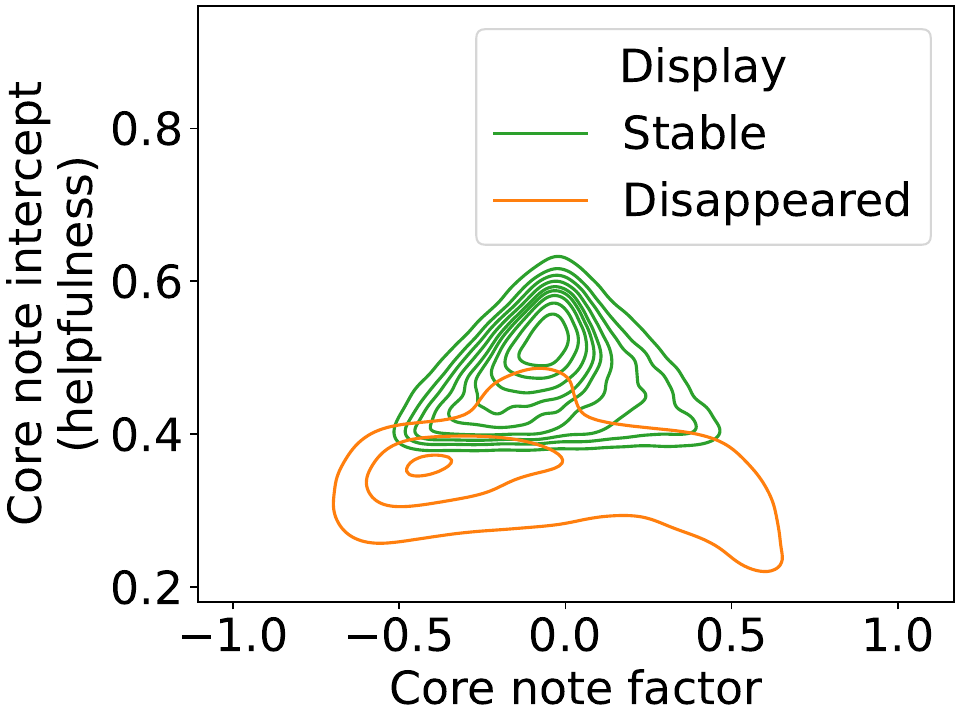}
    \label{fig:note_eval_kde}
    \end{subfigure}
    \hfill
    \begin{subfigure}{0.32\textwidth}
    \caption{}
    \includegraphics[width=\textwidth]{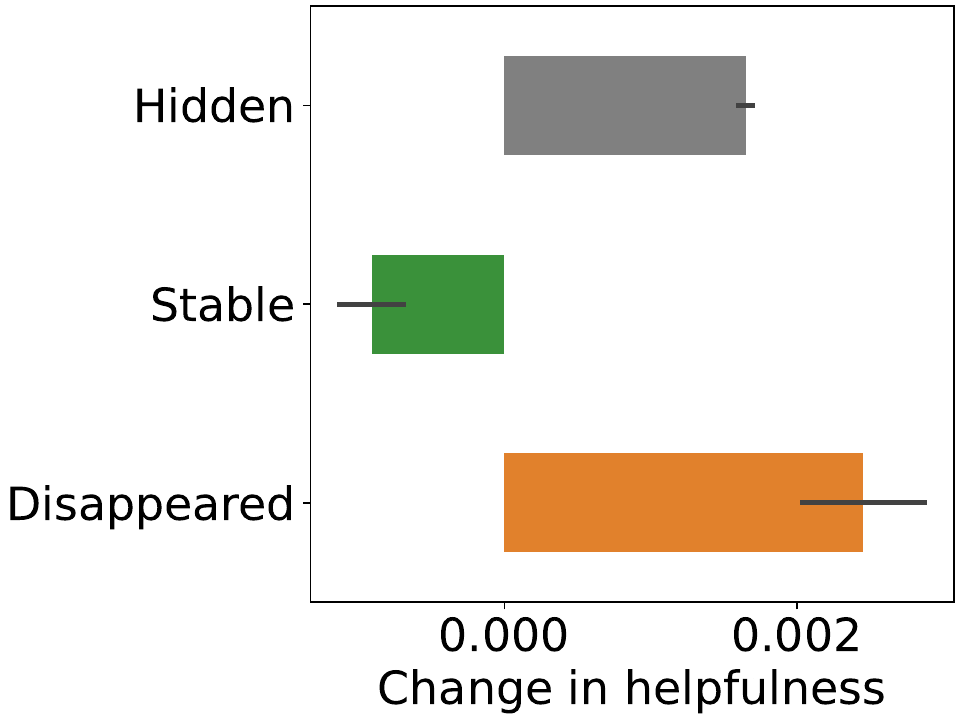}
    \label{fig:score_diff_emb_no_polarized}
    \end{subfigure}
    \hfill
    \begin{subfigure}{0.32\textwidth}
    \caption{}
    \includegraphics[width=\textwidth]{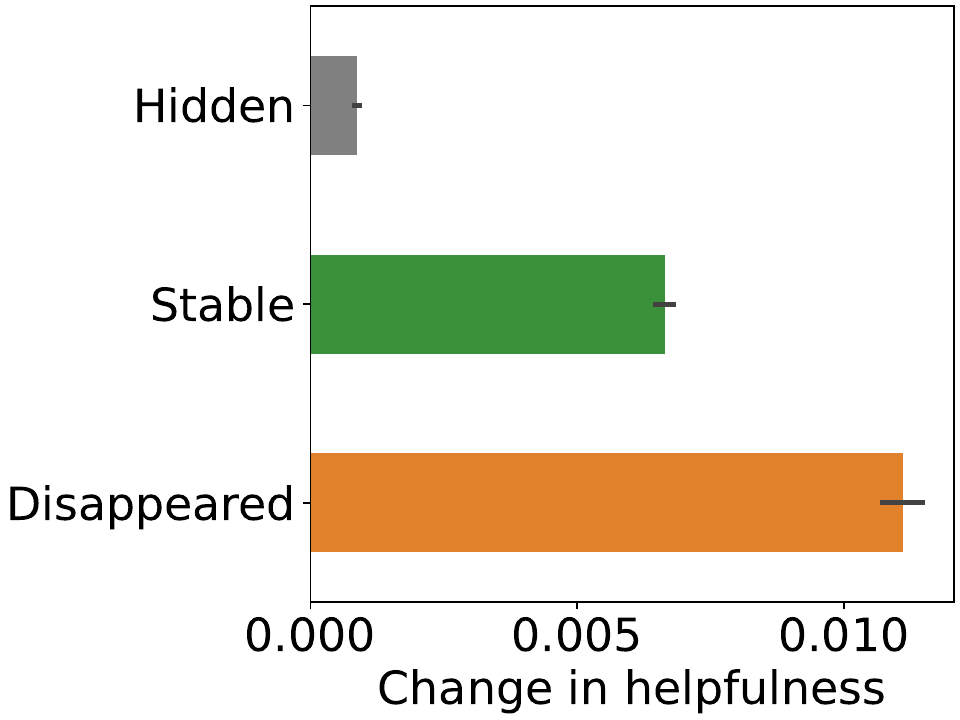}
    \label{fig:score_diff_emb_no_polarized_one_side}
    \end{subfigure}
    \caption{Evaluation of polarized ratings and counterfactual analysis of the note selection algorithm. (a)~Estimated changes in rating leaning for similar and dissimilar raters across levels of political bias in external domains cited in community notes. (b)~Estimated changes in rating leaning for similar and dissimilar raters across high- and low-quality external domains. (c)~Distributions of estimated Core note intercepts (helpfulness) and estimated Core note factors (polarization) from the note selection algorithm, shown for disappeared notes and stable notes. (d)~Changes in helpfulness when ratings from similar and dissimilar raters are excluded. (e)~Changes in helpfulness when ratings from dissimilar raters are excluded. Results are shown for disappeared notes, stable notes, and notes that have never been displayed. The error bars represent 95\% CIs.}
    \label{fig:counterfactual}
    \Description{}
\end{figure*}

\subsection{Counterfactual Analysis}
The observed rating polarization between similar and dissimilar raters may have two competing implications: (i) it may prevent the continued display of potentially problematic or low-quality notes through cross-viewpoint disagreement, or (ii) it may reflect coordinated or ideologically motivated rating behaviors that undermine consensus-based evaluation. Given that the overall rating trend for displayed community notes is strongly positive (Fig.~\ref{fig:general_rating_leaning}), the latter explanation, \ie, ideologically driven rating behaviors, appears more plausible. To further validate this, we assess the political bias and information quality of external domains cited in community notes (see Suppl.~\ref{supp:domain_bias_quality}) and conduct counterfactual analyses by re-running the note selection algorithm after excluding potentially polarized ratings (\ie, those from similar and dissimilar raters).

As shown in Fig.~\ref{fig:domain_bias_rating_leaning}, similar raters exhibit consistent positive rating leanings for community notes with left-leaning ($\var{coef.}=$ \num{0.158}, $p<$ \num{0.001}; 95\% CI: [\num{0.123}, \num{0.193}]), unbiased ($\var{coef.}=$ \num{0.157}, $p<$ \num{0.001}; 95\% CI: [\num{0.115}, \num{0.200}]), and right-leaning domains ($\var{coef.}=$ \num{0.200}, $p<$ \num{0.001}; 95\% CI: [\num{0.113}, \num{0.288}]). In contrast, dissimilar raters show negative rating leanings for notes with left-leaning ($\var{coef.}=$ \num{-0.050}, $p<$ \num{0.001}; 95\% CI: [\num{-0.060}, \num{-0.040}]) and unbiased domains ($\var{coef.}=$ \num{-0.043}, $p<$ \num{0.001}; 95\% CI: [\num{-0.062}, \num{-0.025}]), while their evaluations of notes citing right-leaning domains are not statistically distinguishable from zero. 
A similar pattern emerges when examining domain quality (Fig.~\ref{fig:domain_quality_rating_leaning}): similar raters show positive rating leaning for notes citing both low-quality ($\var{coef.}=$ \num{0.155}, $p<$ \num{0.001}; 95\% CI: [\num{0.039}, \num{0.270}]) and high-quality domains ($\var{coef.}=$ \num{0.161}, $p<$ \num{0.001}; 95\% CI: [\num{0.136}, \num{0.185}]); conversely, dissimilar raters show negative rating leaning, especially for notes with high domain quality ($\var{coef.}=$ \num{-0.047}, $p<$ \num{0.001}; 95\% CI: [\num{-0.056}, \num{-0.038}]). Together, these findings suggest the presence of polarized rating behaviors, \eg, towards notes citing left-leaning or high-quality domains.

Fig.~\ref{fig:note_eval_kde} further shows a clear separation between disappeared and stable notes: disappeared notes are associated with lower estimated helpfulness and higher inferred polarization (farther from zero). To assess whether the polarization in rating behavior between similar and dissimilar raters contributes to the disappearance of community notes, we rerun the note selection algorithm based on ratings from only general raters (\ie, excluding ratings originated from similar and dissimilar raters after note display). Compared with the original helpfulness scores (based on all ratings), counterfactual helpfulness scores increase substantially for disappeared notes with a mild decrease for stable notes (Fig.~\ref{fig:score_diff_emb_no_polarized}). This pattern suggests that dissimilar raters reduce consensus for notes that otherwise would appear more helpful.

Because rating patterns of similar raters are largely consistent with those of general raters, while dissimilar raters show divergent and often oppositional evaluations, we further isolate the contribution of dissimilar raters by excluding only their ratings. As shown in Fig.~\ref{fig:score_diff_emb_no_polarized_one_side}, removing dissimilar raters leads to the largest increases in estimated helpfulness for disappeared notes, followed by stable notes. Overall, the counterfactual analysis supports the interpretation that the increased polarization in ratings following note display, especially negative evaluations from a small group of dissimilar raters, significantly reduce cross-group consensus and contribute directly to note disappearance. 

\section{Discussion}
Here, we examine the stability of community notes displayed on \X and the vulnerability of its consensus-based note selection algorithm. We find that a substantial share of notes (30.2\%) that initially meet the platform's display criteria (\ie, were rated as helpful) later disappear. More critically, disappearance is strongly associated with posts on high-stakes topics (\eg, health, politics) and with posts from high-influence authors (\eg, verified accounts with many followers). We also observe an asymmetry by political orientation: notes displayed on posts from left-leaning authors are more likely to disappear compared to those from right-leaning authors.

\textbf{Disappearance mechanism of community notes.}
Two mechanisms may explain the disappearance of community notes: (i) the notes are genuinely unhelpful or inaccurate, or (ii) the display of community notes may amplify ideologically biased or strategically coordinated rating behaviors that push notes below the display threshold after they appear. Prior work demonstrates that the note selection algorithm is generally effective at surfacing high-quality notes, making it unlikely that widespread disappearance is driven primarily by accuracy issues. Our analysis of rating patterns supports this interpretation. After appearing on a post, notes typically receive more upvotes than downvotes---a pattern that persists even after excluding similar and dissimilar raters who may possess systematic biases. However, we identify systematic polarization in rating behavior between similar and dissimilar raters following note display. Similar raters become more supportive, dissimilar raters become less supportive, and this divergence is particularly pronounced for notes that ultimately disappear. Our counterfactual analysis indicates that this post-display polarization in ratings, particularly from dissimilar raters, plays a substantial role in driving note disappearance. Together, our findings indicate that note disappearance is less a reflection of note quality and more a consequence of ideologically polarized or strategically motivated rating behavior.

\textbf{Vulnerability of consensus-based algorithms.}
While the consensus-based note selection algorithm on \X is designed to surface notes that are perceived as helpful by a broad and ideologically diverse set of raters, emerging evidence suggests that consensus is difficult to achieve: the proportion of notes deemed ``helpful'' and subsequently displayed relative to all community notes over time remains low (around 10\%) and is even declining~\cite{chuai2024community.new,arjmandi2025threats,bouchaud2025algorithmic,razuvayevskaya2025timeliness}. Although politically balanced crowds have demonstrated the capacity to identify misinformation effectively at scale~\cite{allen2021scaling,martel2024crowds,saeed2022crowdsourced,he2025survey}, community-based fact-checking systems remain fragile in polarized social media environments, where partisan disagreement, motivated reasoning, and coordinated rating behaviors can destabilize consensus-based moderation~\cite{bouchaud2025algorithmic,truong2025community,kahan2017motivated,kaufman2022who,allen2022birds,kaufman2022who,vinhas2022fact,walter2020fact}. For instance, users are more likely to provide negative ratings to evaluations from counter-partisans~\cite{allen2022birds,shin2017partisan}. Therefore, many factually accurate but politically sensitive notes often fail to meet the cross-group consensus threshold and remain unpublished, reflecting systematic suppression of information that is credible but polarizing~\cite{bouchaud2025algorithmic}. 

Our findings extend this literature by showing that the \emph{display} of community notes itself can intensify polarization in subsequent ratings. Even when only a small fraction of ratings exhibit systematically oppositional or strategically coordinated behavior, their concentrated negative ratings disproportionately contribute to note suppression~\cite{truong2025community}. This dynamic not only undermines the persistence of currently displayed notes but may also reinforce rating biases in subsequent update rounds, further reducing the proportion of notes that successfully reach publication. More broadly, such post-display polarization may undermine the long-term trustworthiness and effectiveness of community-based fact-checking. While displayed notes can increase user trust and engagement with misinformation~\cite{drolsbach2024community,kankham2025community,slaughter2025community,chuai2024community.new}, polarized ratings can cause high-quality notes to disappear, potentially eroding their effectiveness. Furthermore, the scalability and timeliness of Community Notes system rely on efficient aggregation of crowd ratings~\cite{razuvayevskaya2025timeliness,godel2021moderating,jia2024collaboration}, yet biased or coordinated post-display ratings threaten this efficiency. Our findings therefore highlight the importance of mitigating post-display polarization in rating behavior to ensure the efficacy and sustainability of community-based fact-checking systems.

\textbf{Practical implications.}
Our study underscores the importance of carefully designing post-display update mechanisms and interventions, as display itself can amplify polarization in subsequent evaluations. This has several practical implications for platforms relying on consensus-based moderation and community-based fact-checking. First, platforms may incorporate safeguards to detect and exclude potentially coordinated or manipulated rating behaviors, thereby improving the resilience of consensus-based algorithms against bias and manipulation~\cite{cn2025safeguards}. Second, platforms could consider adopting a multi-layer fact-checking ecosystem that incorporates contributions from both crowd and expert fact-checkers, with experts taking the lead on high-stakes or politically charged content~\cite{dailey2014journalists,zhao2023insights,he2025survey,augenstein2025community}. Such hybrid arrangements may enhance accuracy, reduce vulnerability to polarized rating behaviors, and ensure more stable fact-check visibility on social media platforms. Finally, advancements in Large Language Models (LLMs) offer promising opportunities to support and strengthen community-based fact-checking~\cite{li2025scaling,hassan2019examining,augenstein2025community}. For instance, LLMs can assist contributors with evidence retrieval, note generation, and information synthesis~\cite {de2025supernotes,zhang2025commenotes,wu2025beyond,li2025scaling,hassan2019examining}, and may also help reduce bias in rating processes by providing structured evaluation assistance~\cite{mohammadi2025ai}.

\textbf{Limitations and future work.}
Our results point to several promising directions for future research. First, while we document substantial post-display polarization in rating dynamics, the current data do not permit a clear separation between individual bias, partisan disagreement, and coordinated manipulation. Future studies combining behavioral traces with survey or experimental data could clarify the mechanisms underlying polarized evaluations of community fact-checks. Second, the generalizability of our findings is conditioned by the specific design features of Community Notes on \X, especially the platform's evolving criteria for note display. As other platforms experiment with community-based fact-checking, comparative analyses could identify which design elements enhance the stability of evaluations and protect against coordinated manipulation. Finally, our study relies on a snapshot of the Community Notes dataset, which may omit notes or ratings that were later removed or edited. Longer-term archival efforts that provide more complete historical traces would enable richer analyses of the lifecycle of notes and how contributor behavior evolves over time.


\begin{acks}
This research is supported by the Luxembourg National Research Fund (FNR) and Belgian National Fund for Scientific Research (FNRS), as part of the project REgulatory Solutions to MitigatE DISinformation (REMEDIS), ref. INTER\_FNRS\_21\_16554939\_REMEDIS. Furthermore, this research is supported by a research grant from the German Research Foundation (DFG grant 492310022).
\end{acks}

\newpage
\bibliographystyle{ACM-Reference-Format}
\bibliography{refs}

\newpage
\appendix

\begin{center}
    \Large \textbf{Supplementary Materials for \\ ``Consensus Stability of Community Notes on X''}
\end{center}

\renewcommand\thetable{S\arabic{table}}
\setcounter{table}{0}
\renewcommand\thefigure{S\arabic{figure}}
\setcounter{figure}{0}
\renewcommand\thesection{S\arabic{section}}
\setcounter{section}{0}

\vspace{3mm}
\section{Status Updates on Community Notes}
\label{supp:note_status}

When a community note is first submitted, it is assigned the status of ``Needs More Ratings'' (see Fig.\ref{fig:note_disappear_framework} in the main paper). When community notes receive enough ratings from raters with a broad range of perspectives, their statuses will be updated to either ``Currently Rated Helpful'' or ``Currently Rated Not Helpful'' by the note selection algorithm \cite{x2024algorithm}. Then, the note selection algorithm updates the note status regularly by incorporating newly submitted ratings. Therefore, community notes that have received helpful status and been displayed on the misleading posts can return to ``Needs More Ratings'' status or change to ``Currently Rated Not Helpful,'' thereby losing their display status.

\section{Reproduction of Note Selection Algorithm}
\label{supp:note_selection_algorithm}

\begin{figure}[H]
    \centering
    \includegraphics[width=.8\linewidth]{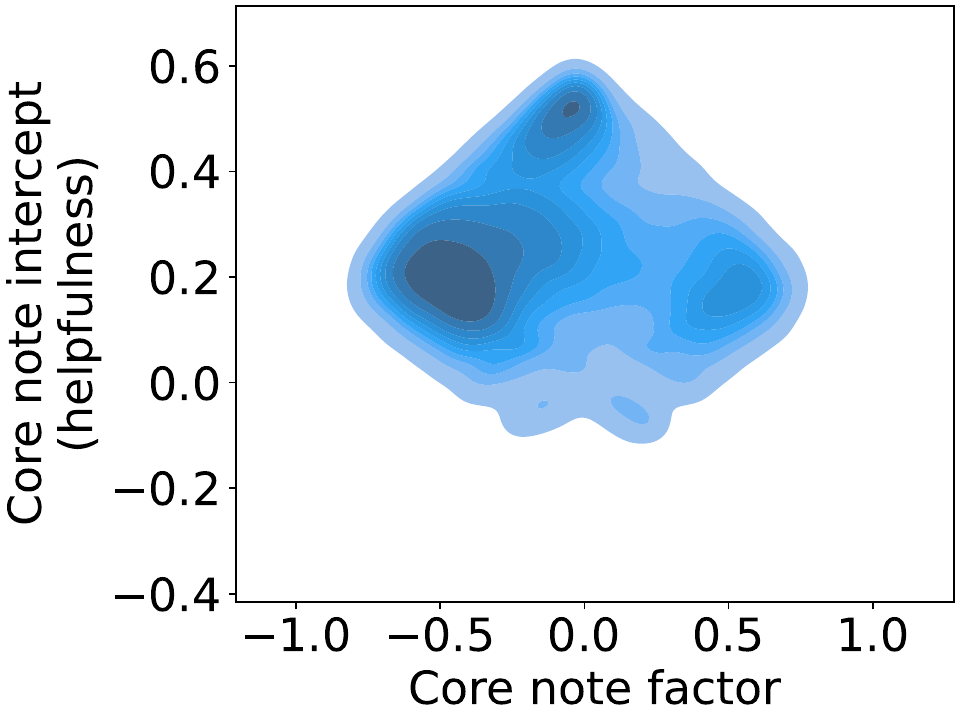}
    \caption{The distribution of estimated Core note intercepts (helpfulness) and estimated Core note factors (polarization) from the note selection algorithm.}
    \label{fig:note_eval_original}
    \Description{}
\end{figure}

The distribution of estimated Core note intercepts (helpfulness) and estimated Core note factors (polarization) from the note selection algorithm is shown in Fig.~\ref{fig:note_eval_original}. To ensure the reliability of our reproduction, we compare the note status generated by the reproduced algorithm with the actual production status. Each community note can receive one of the three statuses: Currently Rated Helpful, Needs More Ratings, and Currently Rated Not Helpful. Overall, 99.8\% of notes in our replication match their corresponding production statuses, highlighting the reliability of our reproduction.

\begin{figure*}
    \centering
    \begin{subfigure}{0.32\textwidth}
    \caption{}
    \includegraphics[width=\textwidth]{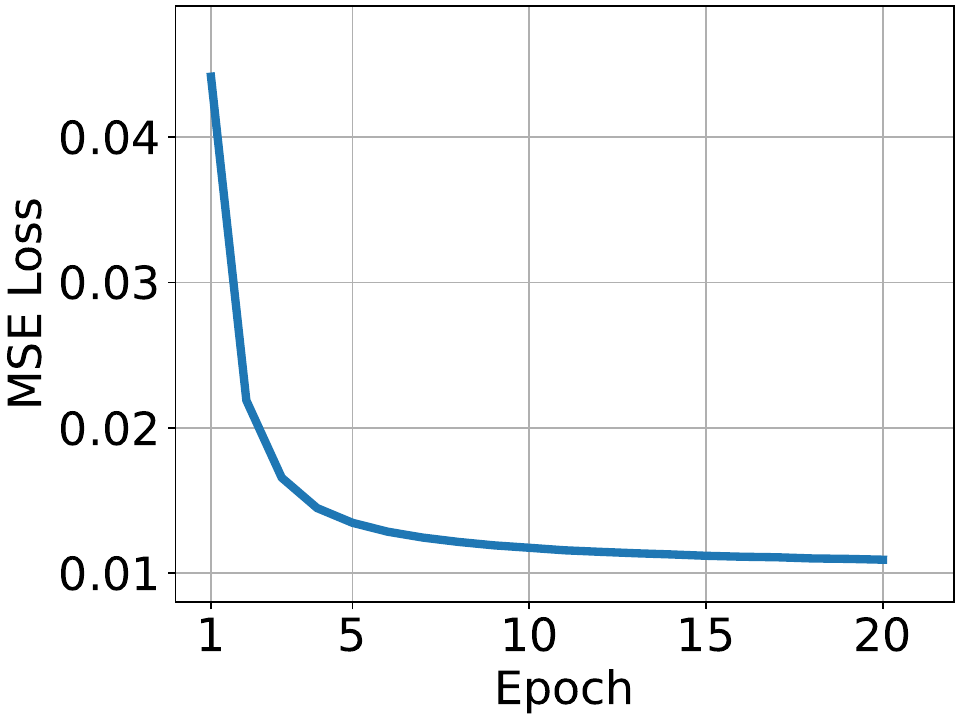}
    \label{fig:mseloss}
    \end{subfigure}
    \hfill
    \begin{subfigure}{0.32\textwidth}
    \caption{}
    \includegraphics[width=\textwidth]{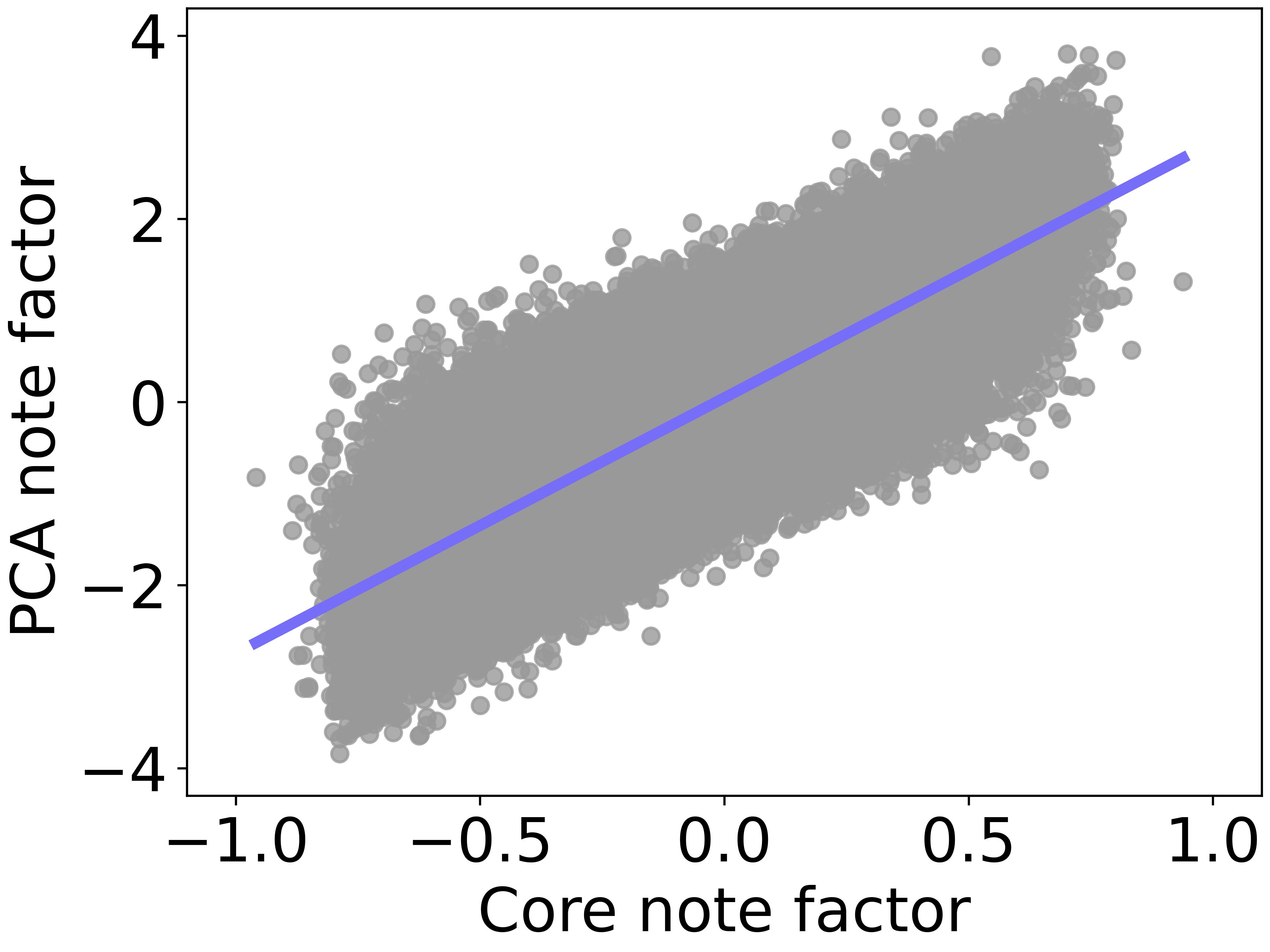}
    \label{fig:note_factor}
    \end{subfigure}
    \hfill
    \begin{subfigure}{0.32\textwidth}
    \caption{}
    \includegraphics[width=\textwidth]{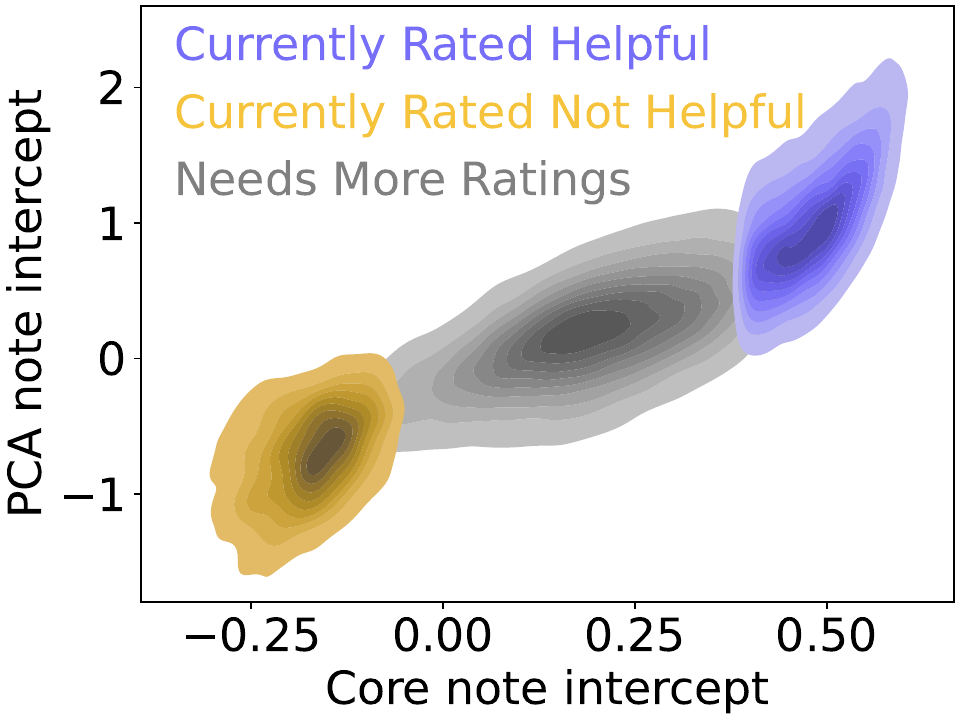}
    \label{fig:note_score}
    \end{subfigure}

    \begin{subfigure}{0.32\textwidth}
    \caption{}
    \includegraphics[width=\textwidth]{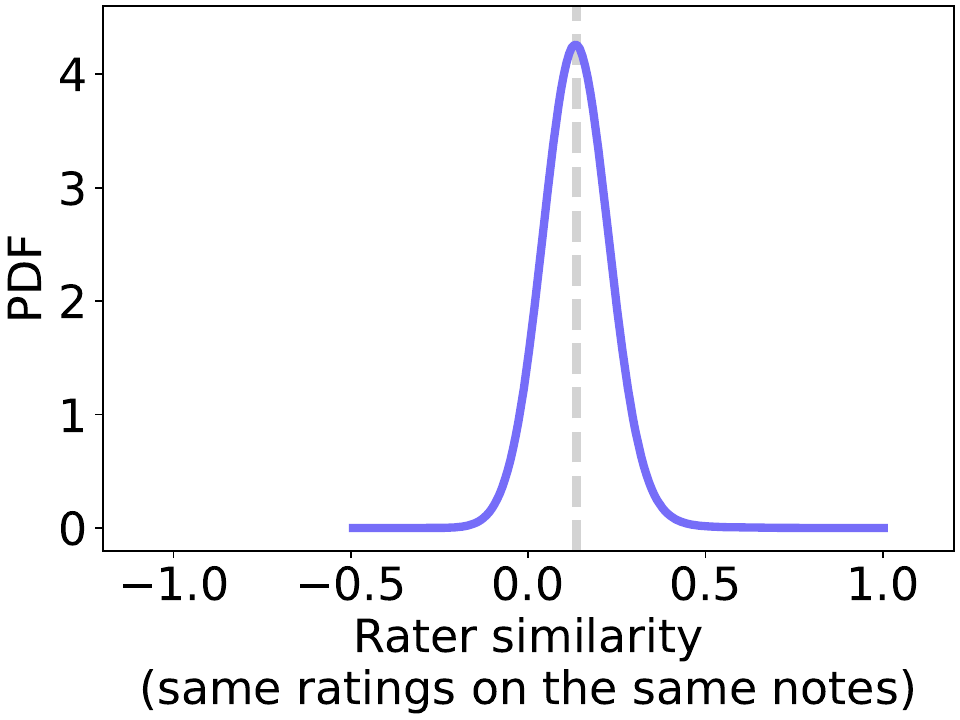}
    \label{fig:cosine_dist_agree}
    \end{subfigure}
    \hfill
    \begin{subfigure}{0.32\textwidth}
    \caption{}
    \includegraphics[width=\textwidth]{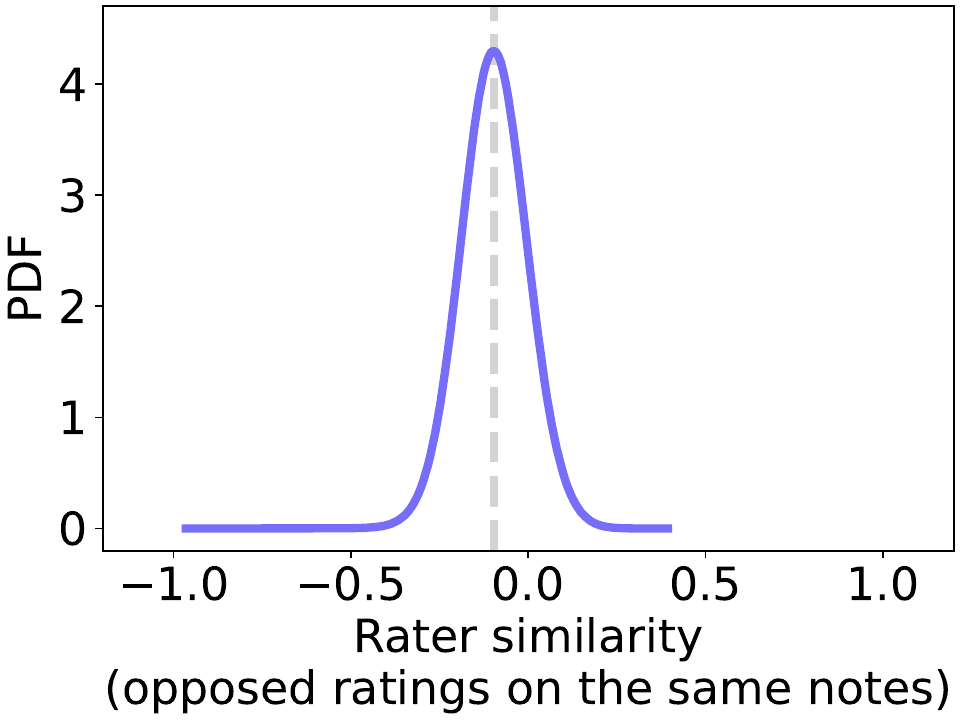}
    \label{fig:cosine_dist_disagree}
    \end{subfigure}
    \hfill
    \begin{subfigure}{0.32\textwidth}
    \caption{}
    \includegraphics[width=\textwidth]{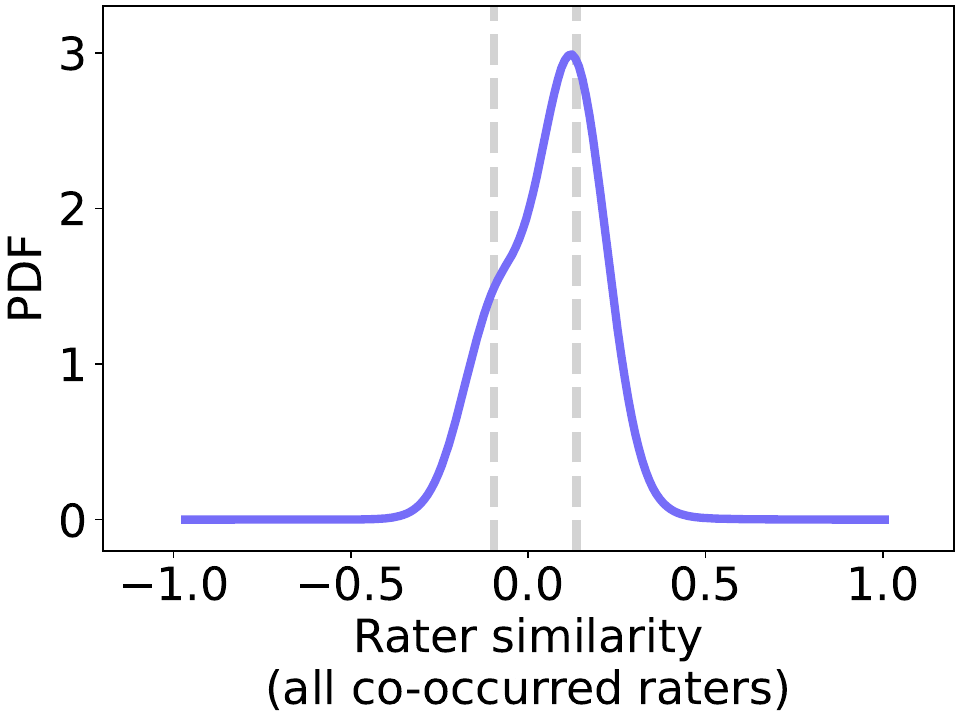}
    \label{fig:cosine_dist}
    \end{subfigure}
    \caption{Analysis of MF-based model for rater representations. (a)~The MSE losses over 20 epochs of training. (b)~The distribution of estimated Core note factors from the note selection algorithm and estimated PCA note factors from the MF-based model for rater representations. (c)~The distribution of estimated Core note intercepts from the note selection algorithm and estimated note intercepts from the MF-based model for rater representations. (d)~The distribution of rater similarity for the same ratings on the same notes. (e)~The distribution of rater similarity for the opposed ratings on the same notes. (f)~The overall distribution of rater similarity across all co-rating pairs.}
    \label{fig:note_eval}
    \Description{}
\end{figure*}

\begin{figure*}[t]
    \centering
    \begin{subfigure}{\textwidth}
    \caption{}
    \includegraphics[width=\textwidth]{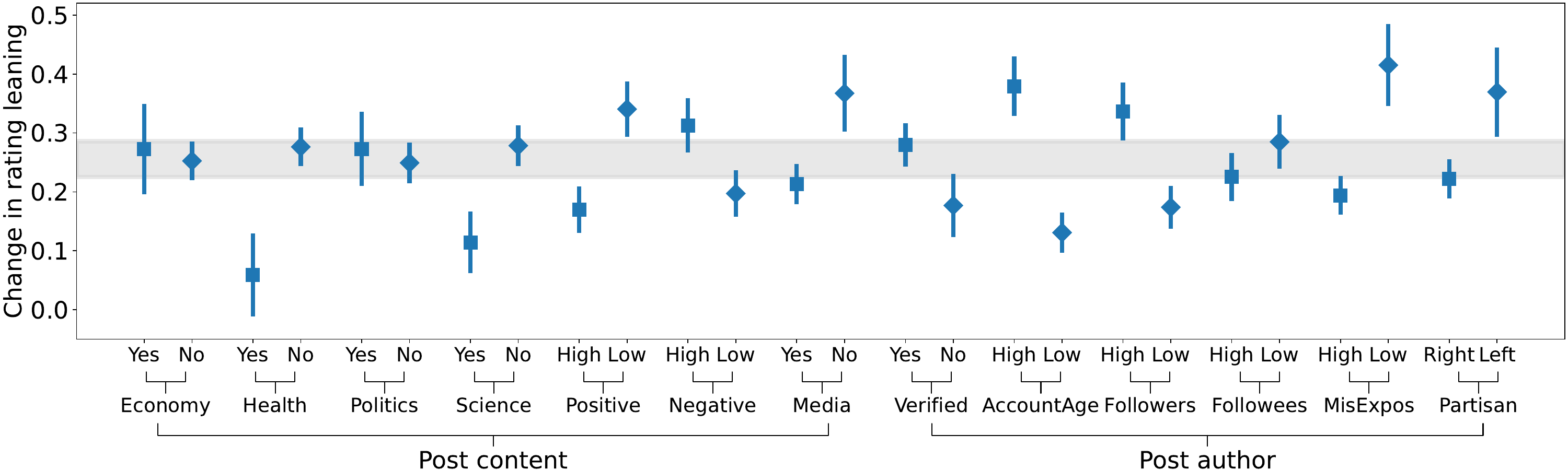}
    \label{fig:leaning_effect_sensitivity}
    \end{subfigure}

    \begin{subfigure}{\textwidth}
    \caption{}
    \includegraphics[width=\textwidth]{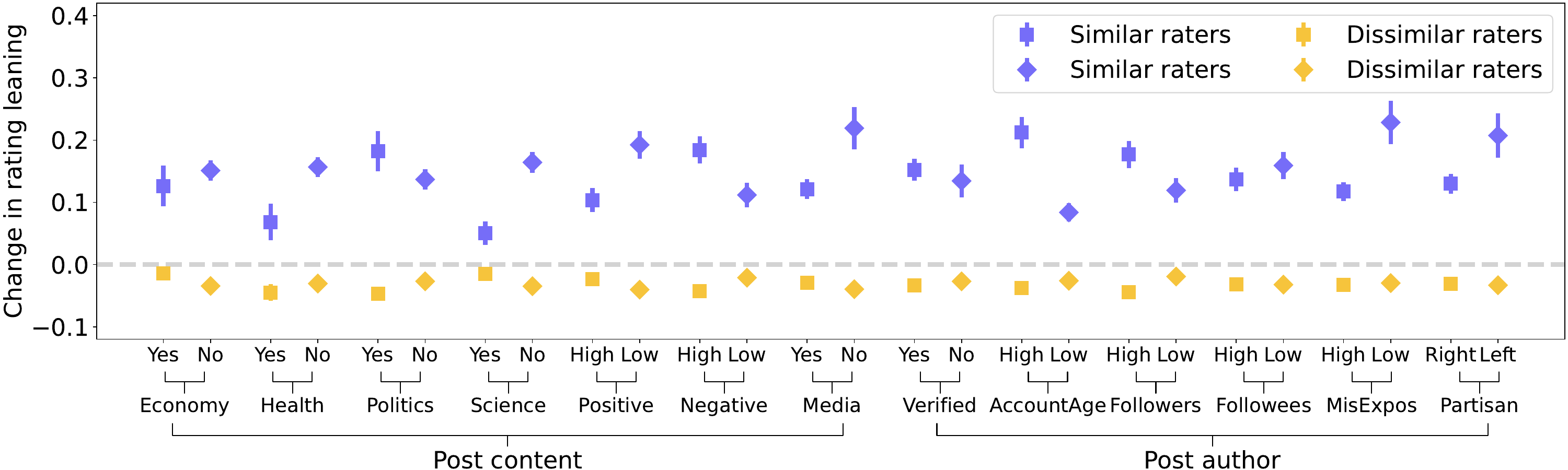}
    \label{fig:polarized_leaning_sensitivity}
    \end{subfigure}
    \caption{Sensitivity analysis. (a)~Estimated changes in rating leaning for all displayed notes after initial display, across different post content and post author characteristics. The error bars represent 95\% CIs. (b)~Estimated changes in rating leaning for similar and dissimilar raters, across different post content and post author characteristics. The error bars represent 95\% CIs.}
    \label{fig:sensitivity_analysis}
    \Description{}
\end{figure*}

\section{Matrix Factorization for Rater Similarity}
\label{supp:mf}

The visualization of MSE training losses of adapted MF algorithm, its comparison with original note selection algorithm, and the distributions of rater similarity across different groups are shown in Fig.~\ref{fig:note_eval}.

\textbf{Validation of the model estimates.}
We validate the reliability of the parameter estimates from the adapted MF model by comparing them with the corresponding parameters estimated by the note selection algorithm: note factors and note intercepts. Given that the note factors in the adapted MF model are 200-dimensional vectors, we apply PCA to reduce their dimensionality for comparison with one-dimensional Core note factors in the note selection algorithm. The Core note factors and the PCA note factors have a significant positive linear relationship ($\var{coef.}=$ \num{2.790}, $p<$ \num{0.001}) with $R^{2}=$ \num{0.792} (see Fig.~\ref{fig:note_factor}, Suppl.~\ref{supp:mf}). Additionally, we examine the distributions of estimated note intercepts, \ie, indicator of helpfulness, in the note selection models and the adapted MF model. We find three clearly separated clusters indicating notes with the statuses of Currently Rated Helpful, Needs More Ratings, and Currently Rated Not Helpful (see Fig.~\ref{fig:note_score}). This suggests that the adapted MF model produces note factors and intercepts that are highly consistent with the estimates from the note selection algorithm, effectively preserving the distinctions among different note statuses.

\section{Characteristics of Posts and Post Authors}
\label{supp:post_characteristics}

We consider various characteristics of posts and their authors, which could influence both the spread of misleading posts and the evaluation of associated community notes.

\noindent \underline{Post characteristics.}
\begin{itemize}[leftmargin=*]
    \item \textit{Topics}. The dataset includes four topic categories---\ie, Politics, Science, Health, and Economy. Each post may be associated with multiple topics, and each topic category is represented as a binary variable.
    \item \textit{Sentiments}. Each post is annotated with positive and negative sentiment scores, represented as continuous variables that capture the polarity of emotional expression in the content.
    \item \textit{Media}. A binary variable indicating whether the post contain media elements ($=$ 1) or not ($=$ 0).
\end{itemize}

\noindent \underline{Post author characteristics.}
\begin{itemize}[leftmargin=*]
    \item \textit{Verified}. A binary variable indicating whether a given post account is verified ($=$ 1) or not ($=$ 0).
    \item \textit{Account age}. A continuous variable indicating the account age (in days) of a given post account since its creation.
    \item \textit{Followers}. A count variable indicating the number of followers of a give post account.
    \item \textit{Followees}. A count variable indicating the number of followees of a give post account.
    \item \textit{Misinformation exposure score and partisan score}. Following the approach proposed by~\citet{mosleh2022measuring}, the misinformation exposure score of an author, ranging from 0 to 1, represents the proportion of followed public figures whose statements were rated as false by PolitiFact. The partisan score, ranging from \num{-1} (Democrat) to \num{+1} (Republican), captures the ideological leaning of an author based on the political orientation of followed public figures.
\end{itemize}

\section{Sensitivity Analysis}
\label{supp:sensitivity}

The estimation results for our sensitivity analysis are shown in Fig.~\ref{fig:sensitivity_analysis}. Fig.~\ref{fig:leaning_effect_sensitivity} presents estimated changes in rating leaning for all displayed notes after initial display, across different post content and post author characteristics. Fig.~\ref{fig:polarized_leaning_sensitivity} presents estimated changes in rating leaning for similar and dissimilar raters, across different post content and post author characteristics.

\section{Evaluation of Domain Bias and Quality}
\label{supp:domain_bias_quality}
We assess the political bias of external domains in community notes using the media bias dataset from Media Bias/Fact Check---the most comprehensive media bias resource on the internet (\url{https://mediabiasfactcheck.com}). Specifically, the domains in the categories of ``Least Biased'' or ``Pro-Science'' are coded as unbiased ($=$ 0). The domains in the categories of ``Left'' ($=$ \num{-1}) or ``Left-Center'' ($=$ \num{-0.5}) are coded as left-leaning. The domains the categories of ``Right'' ($=$ \num{1}) or ``Right-Center'' ($=$ \num{0.5}) are coded as right-leaning. When a note cites multiple domains, we average their bias scores, with positive values indicating right-leaning and negative values indicating left-leaning. Additionally, we measure the information quality of domains in community notes based on a large-scale domain-quality dataset compiled by~\citet{lin2023high}. The rating score ranges from 0 to 1, with a value of 0.5 or higher indicating high quality. We average the ratings for notes containing multiple domains.

\end{document}